# Measuring device suitable for linear distances

Zaid A. I. Alsmadi[a*], Ahmad B. B.Badry[a], Irfan A. Badruddin[1], T. M. Indra Mahlia[a]

[a] Department of mechanical Engineering
faculty of Engineering
University of Malaya
50603 Kuala Lumpur, MALAYSIA

\* Corresponding author**.**

Postal address: Department of mechanical Engineering, faculty of Engineering, University of Malaya, 50603 Kuala Lumpur, MALAYSIA

Email: zaes75@yahoo.com

Tel: 00601-6308-5084

Fax: 00603-7957-5317

**Abstract :**

Measuring device is proposed for determining a linear dimension. The device comprises three associated longitudinally moving parts one of which is a scale. The integer part of the device reading is being taken from the standard millimeter or inches scale , and The fine measurement ( smaller than the minimum scale division) is being done by a setup of two sliders coupled to the device. The first slider includes measuring points. And the other slider includes measuring line. The decimal part of the reading is being taken in such a way that the measuring points reading is related to the fractions of the displacement between the graduated scale and the corresponding measuring line.

1. **Introduction:**

Measuring devices comprising elongated and relatively slidable members with a cursor moveable along the device are well known in the measuring arts. The documents cited hereafter are merely a selection which disclose the general idea behind this group on inventions ; US6289598, GB865285A, US6289598B1, GB184168A, GB1153846A [1-5]. For more details, a graphical comparison and analysis for the ideas included in the inventions abovementioned is available in appendix 1 and 2. However, invention requires cursors wherein a sighting or indicating lines is disposed thereon at an acute angle to the generally longitudinal extent of the device is not common feature.

In this report a new measuring device is proposed for determining a linear dimension. The device comprises three associated longitudinally moving parts one of which is a scale. The integer part of the device reading is being taken from the standard millimeter or inches scale , and The fine measurement ( smaller than the minimum scale division) is being done by a setup of two sliders coupled to the device. The first slider includes measuring points. And the other slider includes measuring line. The decimal part of the reading is being taken in such a way that the measuring points reading is related to the fractions of the displacement between the graduated scale and the corresponding measuring line.

**Appendix C: The device's idea:**

1. Consider a one dimensional measurement scale like the one in figure (8), the scale consists of zero-dimension points, and the distance between the grades of the scale are limited and fixed to $\Delta$.

2. If an extra dimension is added, the scale changes into a $2D$ surface, instead of being a line. And each point on the original scale will stretched to a line. It is proven in appendix 3 that for cubes of dimentions $\geq 2D$ the difference in hypotenuse length between the $(n-1)D$ cube and the $n$ cube decreases with increasing $n$ and vanish when $n$ approach $\infty$, with maximum ratio equals to $\sqrt{2}$ at $2D$. the target of this report is to establish and use the $2D$ surface of the standard ruler to enhance the readability and accuracy of the scale.

3. Consider the $1D$ scale divisions engraved on the surface of the ruler that is line $OA_n$ In figure(1). Starting from the zero point $O$ Assign to every division point a number $A_n$ Where $n$ is the division point number. And the distance between any two points $A_m$ And $A_n$ Is given by $A_m - A_n = \Delta(m - n)$. Stretch the division points along the surface of the ruler to line $B$, as in figure (9), so that every division point $A$ will have a B-image. The length of the lines $\overline{A_n B_n}$ Or simply $\overline{AB}$ Is fixed all over the scale.

4. In order to measure distances using the setup aforementioned, the follow procedure is proposed :

    a. Draw the hypotenuse between any division point $A_{n-1}$ And point $B_n$, as in figure (10). The length of this hypotenuse is $\sqrt{\Delta^2 + \overline{AB}^2} = H$ and it makes an angle $\theta$ with the scale line. The projection of $H$ on the scale-line direction is equal to $H cos\theta = \Delta$

    b. By maintaining the length and the slope of $H$, it is possible to slide it literally along the whole scale as in figure (11). Notice that every time the lower end of the line $A'$

coincides a graduation point the other end $B'$ coincides the B-image of the next graduation point. This is equivalent to $A_{n-1}A' = B_nB'$ .and the following equation is correct

$$\Delta = \text{projection of } A'x \text{ on the scale line } + \text{ projection of } xB' \text{ On the B\_image line}$$

$$= A_{n-1}A' + A'A_n = B_nB' + B'B_{n+1} = A'x\cos\theta + xB'\cos\theta$$

5. The overall distance from the zero point O to the *measuring point* $A'$ Is sketched in figure (12) and is given by the equations

$$OA' = n\Delta + \overline{B_nB'} \quad = (n+1)\Delta - \overline{A'A_{n+1}}$$

$$= n\Delta + \overline{B'x}\cos\theta = (n+1)\Delta - \overline{A'x}\cos\theta$$

As have been shown, the current theory targets to establish a relation between the standard $1D$ scale, and the extension of the graduation points along the device surface ($2D$) surface. This target is achieved by knowing the slope and length of line $H$, and determining its intersection point with the scale graduation points. The aforementioned idea, is suitable to measure distances, and easily it is possible to use it as a base for a simple measurement device.

Theoretically lines are one-dimensional entities, they do not have thickness. Practically, lines posses a specific thicknesses, and As the thickness increased the certainty in measurement thus decreased. However, it is a main objective of this report to design a measuring device to determine precisely the intersection points of two lines ,or equivalently, between a line and a point as well be shown later. To overcome this obstacle, a set up compromising a slider with one *measuring line*, movable along the device surface is introduced to substitute the numerous lines on the surface of the device.

**Practical considerations :**

For purely practical purposes, such as the cost of fabrication and the ease of use, the real device will have only one measuring line traced on the surface of a slider. This arrangement allows at every moment of causing the measuring line on the surface of slider (2b) to correspond with the graduation line $A_n$ Traced on the scale. Furthermore, the hypotenuse line will be substituted by a set of point equally separated, along the direction of $A'B'$. The number of points is determined according to the size of the device and the accuracy required. The set of point is called "measuring points". $OA' = n\Delta + \overline{A'A_n}\cos \cong n\Delta + number\ of\ points/10$

2. **Basic components of the device:**

   1. Device's body :
      a. Longitudinaly extended body,
      b. with scale and pair of jaws.

   2. Fine-reading mechanism, which consists of :
      a. Transparent slider containes the measuring points.
      b. Transparent adjustable-slider contains the measuring line and two Coinciding grides.
      c. A releasably-lockable pin to control slider (a).
      d. Trapizoidal slider to caliprate the slider (b).
      e. Screw with rotating disk.
      f. Return spring.

### 2.1. The body of the device :

Along the width of the device a graduated scale (1a) , in inches or millimeter, is traced In a way that can be related to its length. and a pair of jaws (1b) for engaging an object to be measured is attached to the body, One of them is mounted on a slider which can slide along the scale so that this latter jaw is adjustable. When the two jaws are in contact, the reference line A′B′ traced on the adjustable slider surface coincides with the zero of the graduation of the scale. Figures (1) and (2) shows the basic components of the body.

### 2.2. The sliding parts (Fine-Reading part) :

This part compromises a suitable transparent slider (2a), in which the measuring points are engraved along an inclined line at $\theta$. In figure (1) it is shown that slider could be moved independently of the main scale. the measuring points traced on it's surface are small and equally-spaced dots. On the other hand, the adjustable-Slider (2b) includes a measuring line and two Coinciding grides, the distance between the measuring line and the coinciding grides is always and everywhere equal $\Delta$ .

Parts (2d+2e+2f) in fugure (3) are the component of a fine-tunning mechanism to accurately move the slider (2a) . The trapizoidal slider (2d) is attached to a screw (2e). The rotation of the screw moves the trapizoidal slider downward/upward. In turn the trapizoidal slider (2d) force adjustable-Slider (2b) to move parallel to the scale line. And after taking the correct reading a return spring (2f) is used to send slider (2a) back to its initial position.

### 3. Measuring procedure :

1. Release pin (2c) and make sure that the first measuring point kisses the measuring line by operating screw (2c) in the suitable direction as in figure (2).

2. Firmly engage the body to be measured between the jaws. lock pin (2c) to hold the reading. as in figure (4)

3. From the main scale, record the integer part of the reading n which is the graduation point enclosed between the measuring line, as in figure 5.

4. operate screw (2e) until the two Coinciding grides on the slider is coincided with the two graduation lines $A_{n-1}$ And $A_{n+1}$ Traced on the scale. see figure (6) for illustration.

5. The correct reading is the point which is being intersected by the *measuring line* coinciding the graduation point $A_{n+1}$, precisely and completely as in figure (7).

6. For numerical values let $\Delta = 1mm$ and $\overline{AB} = 10mm$ the value of $\theta = 84.26°$ Based on this values the device's equation reduces to this equation

$$OA' = n\Delta + number\ of\ points/10$$

**Acknowledgment :**

The authors gratefully acknowledge Universiti Malaya Research grant No. RG074-09AET . also the first author would like to thank Dr. Reza Rahbari for useful discussions.

**Appendix D :**

Consider the case of $n$ dimentional cubes in figure (13) where the length of the edge is equal to ($\Delta$). The $1D$ cube has hypotenuse equal to $\sqrt{\Delta^2} = \Delta$. In the $2D$ case the hypotenuse is equal to $\sqrt{\Delta^2 + \Delta^2} = \sqrt{2}\Delta$. In the $3D$ case the hypotenuse is equal to $\sqrt{\Delta^2 + \Delta^2 + \Delta^2} = \sqrt{3}\Delta$. In the $nD$ case the hypotenuse is equal to $\sqrt{\Delta^2 + \cdots + \Delta^2} = \sqrt{n}\Delta$. The ratio of the 2D hypotenuse to the 1D hypotenuse is $\sqrt{2}\Delta/\Delta = \sqrt{2}$. The ratio of the 3D hypotenuse to the 2D hypotenuse is $\sqrt{3}\Delta/\sqrt{2}\Delta = \sqrt{3/2}$. The ratio of the $nD$ hypotenuse to the $(n-1)D$ hypotenuse is $\sqrt{n}\Delta/\sqrt{n-1}\Delta = \sqrt{n/n-1}$. When $n \to \infty$ The ratio $\sqrt{n/n-1} \approx 1$ and the difference in hypotenuse length between the $(n-1)D$ cube and the $n$ cube decreases with increasing $n$ and vanish when $n$ approach $\infty$.

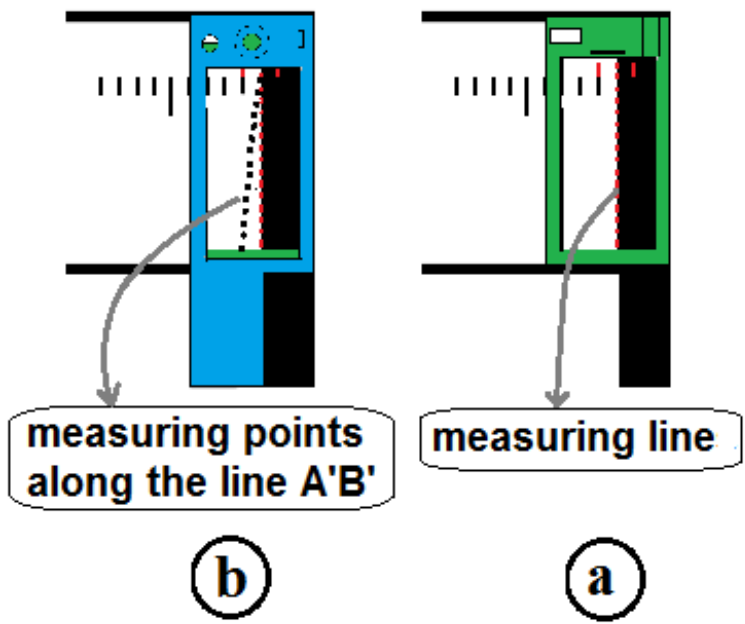

**Figure 2** : the tow jaws are in contact , the *measuring line* coincides with the 1st two lines on the scale, and all the points are enclosed between the two *measuring line*s.

**Figure 1** assembling the scale and the two sliders

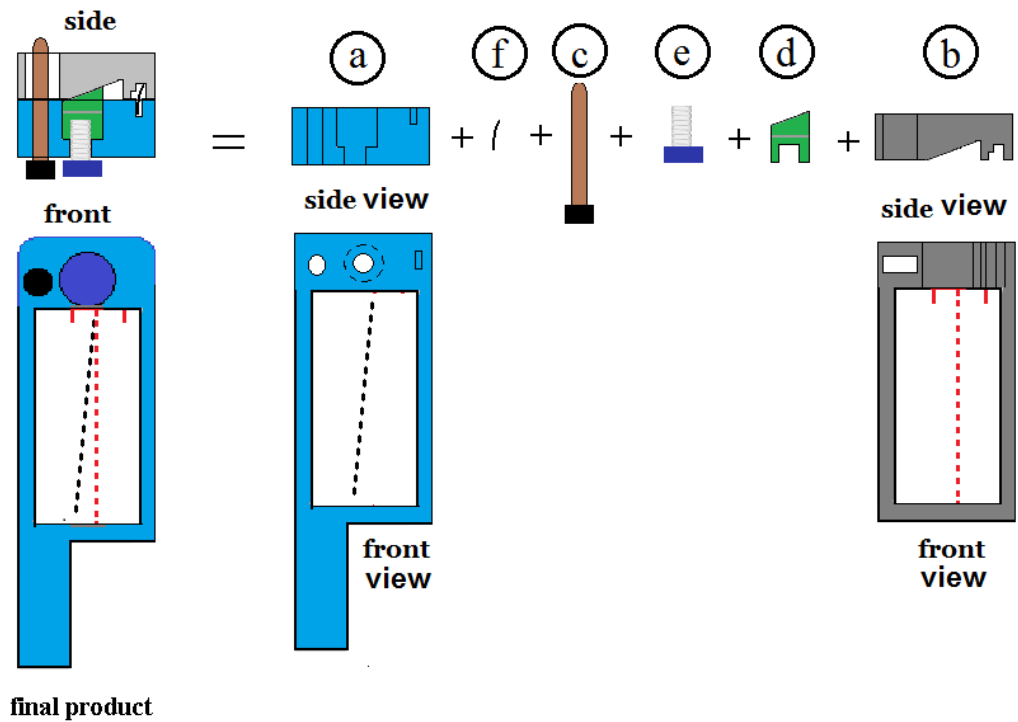

**Figure 3 :** Assembling the Fine-tunning mechanism

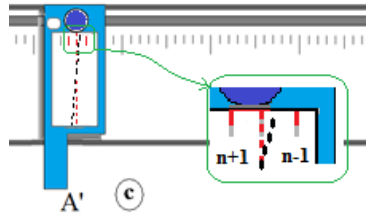

**Figure 4** : the device shown during measuring, in which the tow jaws are attached to the measured body, and all the points are enclosed between the two *measuring line*s, the releasably-lockable pin (c) is used to hold the *measuring-points* position

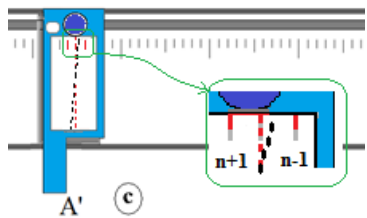

**Figure 6** : reading stage, the releasably-lockable pin holds the *measuring line*s, the screw with rotating disk (e) is used to operate the slider (b), which in turn pushes the *measuring line* until it coincides with adjacent scale grade.

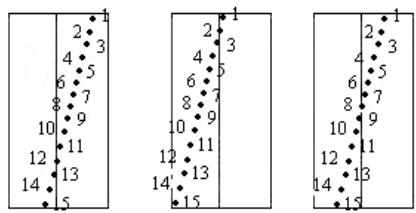

**Figure 7 :** different values of the decimal part of the reading. Notice that the reading is scaled up for the purpose of illustration

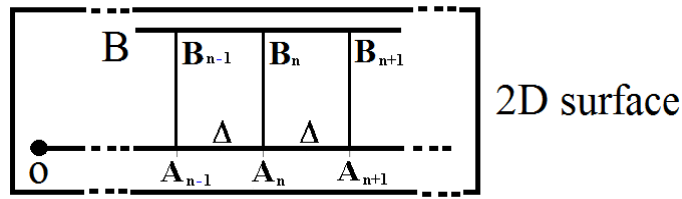

**Figure 9 :** The 1*D* scale One dimensional scale, starts at O and $A_{n-1}$, $A_n$ and $A_{n+1}$ are any three successive points the B-image is at a distance $\overline{AB}$ away on the surface of the device

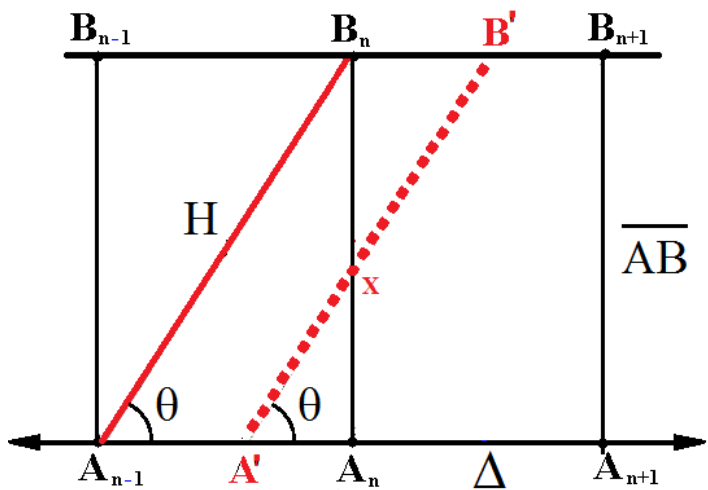

**Figure 10 :** The hypotenuse between division points $A_{n-1}$ on the scale And point $B_n$ on the B-image. then shifting point A' a distance $A'A_n$ shifts point B' equivalent distance in the same direction

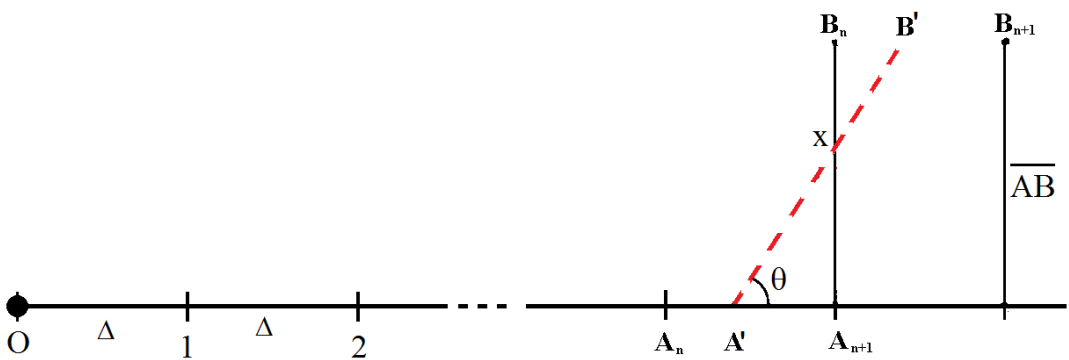

**Figure 12 :** theoretical measuring device compromising a standard scale having equi-distanced graduation points and a set of lines start in the graduation points and extend along the widwth of the 2D scale and a sloped line at $\theta$ slides freely along the scale surface.

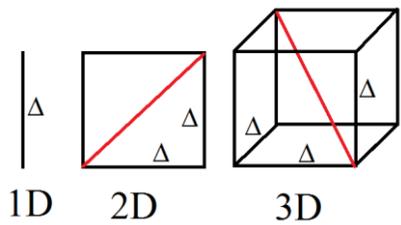

**Figure 13**: multidimensional cubes; the fare left line is a 1D cube, the middle cube is 2D cube and the last cube is a 3D cube